# Laterally Oscillating Trajectory for Undersampling Slices: LOTUS

M. Sothynathan[1,2], P. I. Dubovan[3,4], C. A. Baron[1,5]

[1]Centre for Functional and Metabolic Mapping (CFMM), Robarts Research Institute, Western University, London, Ontario, Canada

[2]Department of Biomedical Engineering, Faculty of Engineering, Western University, London, Ontario, Canada

[3]Athinoula A Martinos Center for Biomedical Imaging, Massachusetts General Hospital, Charlestown, MA, USA

[4]Department of Radiology, Harvard Medical School, Boston, MA, USA

[5]Department of Medical Biophysics, Schulich School of Medicine and Dentistry, Western University, London, Ontario, Canada

**Corresponding author:**

Name: Mayuri Sothynathan

Address: Centre for Functional and Metabolic Mapping

Western University

1151 Richmond St. London, Ontario, Canada N6A 3K7

Email: msothyna@uwo.ca

Tel: 519-697-1570

# 1 ABSTRACT

**Purpose:** While spiral sampling offers SNR advantages for diffusion MRI, its acceleration with simultaneous multislice remains relatively unexplored. This study introduces Laterally Oscillating Trajectory for Undersampling Slices (LOTUS), which is a 3D spiral-like k-space trajectory that aims to minimize g-factor via controlled incoherent aliasing. To aid in validation, we also introduce a robust method to estimate g-factor for iterative non-Cartesian reconstructions.

**Methods:** Simulated data sampling of a numerical phantom was performed using LOTUS and several acquisition schemes proposed by others to quantitatively compare the resulting image quality when compared to a known ground truth. Diffusion-weighted *in vivo* brain data from two subjects was acquired with two in-plane acceleration factors (2x and 4x), and two slice acceleration factors (2x and 4x). Estimated g-factor maps and fractional anisotropy maps were calculated to quantitatively and qualitatively compare trajectory performance. For both simulation and *in vivo*, reconstructions both with and without compressed sensing were utilized.

**Results:** Simulations generally showed decreased g-factor (20%-31%, depending on trajectory, at highest undersampling rate) and improved reconstruction accuracy (mean-square error, structural similarity index, and entropy metrics) for LOTUS compared to the other trajectories. The *in vivo* acquisitions demonstrated g-factor benefits and qualitative image quality improvements that mirrored the simulation results. For both simulation and in vivo, improvements for LOTUS increased for higher numbers of simultaneous slices.

**Conclusion:** By enabling higher rates of slice acceleration, LOTUS shows promise for decreasing scan time, which is especially beneficial for diffusion MRI.

# 2 INTRODUCTION

Diffusion magnetic resonance imaging (dMRI) is an imaging modality that characterizes the diffusion properties of water in tissue to provide valuable insights on the microstructure of the brain. However, dMRI is plagued by long scan times due to the long repetition time (TR) required to acquire high-resolution, whole-brain images for many diffusion directions.[1]

Typically, dMRI acquisitions are performed using single-shot two-dimensional (2D) Echo Planar Imaging (EPI). Standard 2D parallel imaging approaches are primarily used to reduce distortions due to $B_0$ inhomogeneity and have a relatively limited impact on scan time for dMRI, where a large portion of scan time is spent on diffusion encoding. However, methods to decrease scan times are needed since new dMRI representations and models require many diffusion directions and high diffusion weightings.[2] Simultaneous multislice (SMS) allows for the acquisition of multiple image slices during a single shot, reducing acquisition time by a factor equivalent to the number of simultaneously acquired slices.[3] Additionally, SMS mitigates signal-to-noise ratio (SNR) penalties compared to standard parallel imaging approaches since the SNR loss from decreasing the total number of samples is perfectly compensated by the SNR gain from performing a "volumetric" acquisition of multiple slices at once. Thus, the only source of SNR loss is through g-factor losses, but this can be greatly ameliorated via controlled aliasing approaches that reduce overlap of aliased signal via radiofrequency (RF) or "blipped CAIPI" strategies. Notably, blipped CAIPI is equivalent to 3D sampling of k-space, where the total span of k-space acquired in the slice direction is given by the inverse of the spacing between simultaneous slices (Fig. 1a).[1,4]

Non-Cartesian spiral sampling (Fig. 1b) allows for SNR gains in the order of 40-80% in dMRI at 3T due to reduced echo times (TE) resulting from center-out sampling.[5] Additionally, spiral sampling provides reduced g-factor compared to EPI[5], approximately isotropic point-spread functions (PSFs)[6], compatibility with navigator and variable-density sampling approaches, and increases spatial encoding efficiency which is valuable for extensive signal preparations required for dMRI.[7] Nevertheless, clinical adaptability of spiral imaging is limited due to artefacts from unwanted field perturbations.[8] Field changes caused by $B_0$ field inhomogeneity may be monitored via $B_0$ mapping, while recent advancements in field monitoring has enabled the detection of eddy current induced field perturbations.[9,10] These fields may be accounted for by using an expanded encoding model during image reconstruction to mitigate artefacts.[7,11,12]

These advancements in field monitoring have increased the feasibility of spiral dMRI and enables the design of new trajectories. Combining SMS with spiral trajectories has the promise to greatly increase SNR efficiency, but the optimal way to apply controlled aliasing is not clear. Non-Cartesian blipped CAIPI approaches[3,4] as well as tilted hexagonal sampling (T-Hex)[13,14] have shown promise to perform SMS with a spiral-like trajectory. Additionally, preliminary investigations have employed concentric spiral-like rings in k-space and the addition of a sinusoidal z-gradient to a multi-shot spiral trajectory, which have both shown potential for non-Cartesian SMS.[15,16] However, questions remain about the choice of oscillation period for the sinusoidal approach and g-factor performance for all approaches.

This study proposes a 3D k-space trajectory inspired by this earlier work,[15,16] but here the period of oscillations is governed in k-space rather than by the z-gradient. We compare our proposed technique, Laterally Oscillating Trajectory for Undersampling Slices (LOTUS) (Fig. 1e), to CAIPI EPI; standard spiral (i.e., no $k_z$ manipulation); "CAIPI-like" spiral (spiral in $k_{xy}$ and uses gradient blips to sample multiple $k_z$ planes, similar to concentric rings and CAIPI EPI); and T-Hex, a trajectory that is spiral in $k_{xy}$ and samples multiple $k_z$ planes using hexagonal grids in planes orthogonal to the read direction.[13] To evaluate the performance of the trajectories, we introduce a new method to compute g-factor for non-Cartesian trajectories that requires no tuning of number of image reconstruction iterations, in contrast to previous approaches. We find that a LOTUS oscillation period based on the Golden Angle, which evenly distributes $k_z$ oscillations throughout the $k_{xy}$ plane, results in minimal noise amplification. Finally, hypothesizing that the pseudo-random nature of LOTUS may be beneficial for compressed sensing (CS), we assess several image fidelity metrics for compressed sensing reconstructions. Overall, LOTUS has lower g-factor noise amplification than other trajectories. For low numbers of simultaneous slices, LOTUS and blipped spiral have comparable performance considering both g-factor and CS performance, but LOTUS exhibits growing advantages over blipped spiral as the number of simultaneous slices increases.

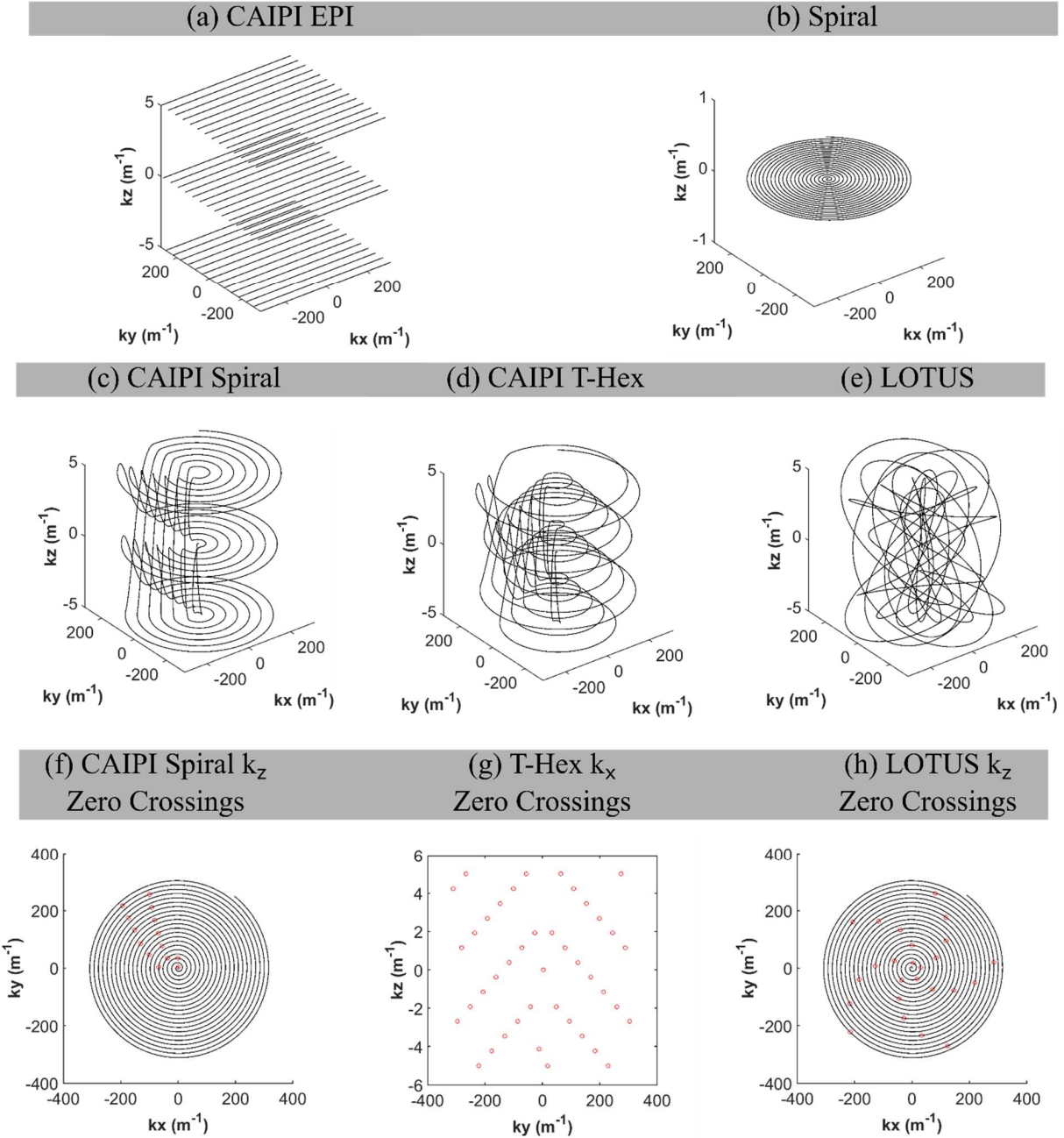

*Figure 1: (a) CAIPI EPI trajectory visualized in 3D, (b) Standard single-shot spiral trajectory (c) "CAIPI-like" spiral trajectory for three simultaneous slices, (d) Tilted Hexagonal trajectory (e) Laterally Oscillating Trajectory for Undersampling Slices (LOTUS) (f) "CAIPI-like" Spiral visualized in x-y plane with $k_z$ zero-crossings indicated in red (g) T-Hex $k_x$ zero crossing visualized in y-z plane (h) LOTUS visualized in x-y plane with $k_z$ zero-crossings indicated in red. For (c) to (h), $R_S = 3$. For (e) and (h), $C_\phi = 0.618$.*

# 3 THEORY

## 3.1 TRAJECTORY DESIGN

Our implementation of the LOTUS trajectory follows the numerical approach developed by Pipe.[17–22] The spiral waveform in 2D k-space ($k_x$, $k_y$) can be denoted in polar coordinates as

$$\overrightarrow{k_{xy}}(t) = k_{xy}(t)e^{i\phi(t)} \quad [1]$$

where $\overrightarrow{k_{xy}}$ represents the k-space vector, and $k_{xy}$ and $\phi$ represents the respective magnitude and phase. For blipped CAIPI, the z-gradient blips create steps in k-space separated by $\Delta k_z = 1/(R_s \Delta z)$, where $R_S$ is the number of simultaneously acquired slices and $\Delta z$ is the slice spacing. Thus, the total $k_z$ coverage over all steps is given by $W_z = \Delta k_z (R_S - 1)$. For LOTUS, we define a sinusoidal oscillation of $k_z$ covering this same span of k-space:

$$k_z = \frac{W_z}{2} \cos(C_\phi \phi) \quad [2]$$

where $C_\phi$ is a user-defined numerical factor that relates the $k_z$ oscillation frequency to the spiral rotation frequency.

The numerical algorithm developed by Pipe creates a discretized trajectory from start to finish one point at a time. For each point, the key equations governing the gradient values are[23]:

$$\overrightarrow{G_{xy}}(t) = G_{xy}(t)e^{i\theta(t)} \quad [3]$$

$$\frac{dk_{xy}}{d\phi} = \frac{\Delta k_{xy}}{2\pi} \quad [4]$$

$$\frac{d\phi}{dt} = \frac{\gamma G_{xy} \sin(\alpha)}{k_{xy}} \quad [5]$$

$$\frac{dk_{xy}}{dt} = \gamma G_{xy} \cos(\alpha) \quad [6]$$

where $G_{xy}$ and $\theta$ are the magnitude and angle of the gradient in the xy-plane, $\Delta k_{xy}$ is the spacing between successive spiral loops (governed by Nyquist criteria and in-plane undersampling rate), and $\alpha = \theta - \phi$ is the angle of the gradient relative to the angle of $\vec{k}(t)$. Taking the derivative of Equation 2 with respect to $\phi$ and multiplying the result with Equation 5 yields:

$$\frac{dk_z}{dt} = \frac{W_z \gamma G_{xy} \sin(\alpha) C_\phi \sin(C_\phi \phi)}{2 k_{xy}} \quad [7]$$

Recognizing that $dk_z/dt = \gamma G_z$, we arrive at the value of $G_z$ relative to the in-plane spiral parameters:

$$G_z(t) = \frac{G_{xy}(t)C_\phi W_z}{2k_{xy}(t)} \sin(\alpha(t)) \sin\left(C_\phi \phi(t)\right) \quad [8]$$

Accordingly, to generate a LOTUS trajectory Equation 8 is used to find G$_z$ for every step in the numerical algorithm (immediately after G$_x$(t) and G$_y$(t) are determined for that step), and constraints for slew rate and maximum gradient consider the net values over all three axes.

Note that it is important to define $k_z$ based on cosine rather than sine so that gradient values are close to zero at the beginning of the trajectory where $t = 0$. However, without intervention this would cause $k_z$ to oscillate between 0 and $W_z$, and thus to center $k_z$ about 0 we apply an additional factor of 0.5 to $G_z(t)$ for $0 \leq C_\phi \phi(t) < \pi$.

## 3.2 G-FACTOR ESTIMATION FOR NON-CARTESIAN MRI

Expanded encoding reconstructions of non-Cartesian data ostensibly solves the equation

$$\hat{x} = argmin_x \|SCx - y\|_2^2 \quad [9]$$

where $x$ is the object-domain image, $y$ is the k-space data, $S$ represents data sampling including B$_0$ inhomogeneity, and $C$ represents receiver sensitivity profiles.[7,24] Iterative reconstructions using the conjugate gradient method are typically used for these problems, but it is a well-known issue that, without regularization, noise will overwhelm the image for single-shot spiral acquisitions. Accordingly, regularization is performed to control noise, either implicitly through early stopping of conjugate gradient iterations[25,26] or other explicit types of regularization.[27] From our observations, the considerable noise amplification for these types of reconstructions stems from the algorithm attempting to fill in the outer unacquired corners of k-space using the high-frequency components of the receiver sensitivity profiles, which is terribly ill-conditioned (Supporting Figure S1). Notably, the noise propagates into the acquired region of k-space as iterations are increased and simply masking the corners of k-space after reconstruction fails to fully compensate for it. These issues present a problem when attempting to characterize g-factor, as the number of iterations strongly affects g-factors and can even lead to $g < 1$ when too low. Here, we propose a simple modification to the above equation to explicitly restrict the reconstruction to only fill in the circular k-space region that has been acquired and avoid extraneous noise amplification:

$$\hat{x} = argmin_x \|SCx - y\|_2^2 + \lambda_k \|M_k F x\|_2^2 \quad [10]$$

where $F$ is a 2D Fourier transform to bring the estimated image to Cartesian k-space, $M_k$ is a mask that selects only the unacquired corners of k-space, and $\lambda_k$ is the strength of suppression of the unacquired parts of k-space. We chose $\lambda_k$ as a very large value that is effectively infinite to completely suppress the unacquired portions of k-space, and the conjugate gradient method was permitted to fully converge.

To assess g-factor related noise amplification, the pseudo-multiple replica method was used.[28] Noise standard deviation maps ($\sigma$) were computed and normalized by the square root of the readout duration to account for differences in acceleration between trajectories and obtain the "normalized noise map", NNM:

$$NNM = \sigma/\sqrt{T_{RO}} \qquad [11]$$

where $T_{RO}$ is readout time. Note that normalization by $\sqrt{T_{RO}}$ is a generalization of the traditional normalization by the square root of the acceleration rate, which is required for non-Cartesian MRI where the acceleration rate is not necessarily an integer value.

For simulations, g-factor maps were computed via

$$g = NNM/NNM_F \qquad [12]$$

where $NNM_F$ is the normalized noise map for a fully sampled reference scan.

For *in vivo* scans, a fully sampled reference scan had a very long readout time leading to excessive T2* decay (~90 ms). Accordingly, here $NNM$ was normalized by the mean NNM (averaged over the entire brain in all slices) computed from $\widehat{NNM}_F$, a scan with no simultaneous multislice and an in-plane undersampling rate of 2, which is expected to have low g-factor noise amplification, to obtain an approximation of g-factor, $\hat{g}$:

$$\hat{g} = \frac{NNM}{\widehat{NNM}_F} \qquad [13]$$

## 3.3 COMPRESSED SENSING RECONSTRUCTION

While pure random sampling is impractical due to hardware and physiological constraints, it is expected that LOTUS will have an incoherent point-spread function (PSF) similar to random sampling. It is hypothesized that this will increase the efficacy of CS, a reconstruction algorithm that increases acquisition efficiency by enabling k-space undersampling. CS solves

$$\hat{x} = argmin_x \frac{1}{2}\|SCx - y\|_2^2 + \lambda\|\psi x\|_1^1 \qquad [14]$$

where $\psi$ is a wavelet transform, and $\lambda$ tunes the CS strength.[29] Notably, the regularization used for CS implicitly suppresses noise from the unacquired corners of k-space, and we have found additional regularization with $\lambda_k\|M_k Fx\|_2^2$ has a negligible effect on the reconstructed image.

# 4 METHODS

## 4.1 TRAJECTORIES

CAIPI-EPI acquires multiple planes of k-space using gradient blips (Fig. 1a). Similar to CAIPI-EPI, A "CAIPI-like" spiral trajectory samples multiple planes of k-space[3,4] (Fig. 1c) and T-Hex[13,14] (Fig. 1d) samples multiple planes of k-space using a hexagonal grid in planes orthogonal to the read direction (Fig. 1g). Meanwhile, LOTUS (Fig. 1e) has sinusoidal $k_z$ with zero-crossings dictated by the period factor $C_\phi$ (Fig. 1h). In experiments, the in-plane acceleration rate, $R_{IP}$, is defined from the separation between successive rings of the spiral, ignoring $k_z$. The slice-acceleration rate, $R_s$, is the number of simultaneous slices.

## 4.2 RECONSTRUCTION

Image reconstruction for both simulations and *in vivo* scans (details below) was performed in MATLAB using the open-source MatMRI toolbox, using an iterative expanded encoding model-based reconstruction.[27,30–32] MatMRI is a graphics processing unit (GPU)-enabled reconstruction framework that can handle any k-space trajectory with the option to include a $B_0$ map, higher-order coefficients fitted from the monitored field dynamics, and implement CS ($\ell_1$-norm regularization). CS image reconstructions of *in vivo* data used an automatically tuned regularization factor[27,29] while a range of regularization factors were explored for simulations because a ground truth image was available for comparison. The delay between the field probe and MRI data was determined using a model-based retrospective algorithm.[33] All experiments were run on a workstation with an Intel i9-11900K processor and 128 GB random-access memory with GPU Nvidia GeForce RTX 4090 with 25.76 GB of GDDR6X memory.

## 4.3 SIMULATIONS

To simulate reconstructions with a known ground truth, a numerical phantom was generated in MATLAB and used as $x$ in the forward encoding model, $y = SCx$, to simulate k-space data. C was created from 16 synthetic receivers with sensitivity decreasing with an inverse-square trend relative to a point approximately a factor of 1.5 outside the FOV. To generate noisy data, complex Gaussian noise was added to the simulated data (SNR for fully sampled spiral ~ 33). Simulations were performed for spiral, CAIPI spiral, T-Hex, and LOTUS acquisitions. A CS reconstruction was performed using MatMRI; however, they did not include a $B_0$ map and second-order or higher coefficients of the expanded encoding model.

### 4.3.1 Investigation of $k_z$ Oscillation Period

To determine the ideal $k_z$ oscillation period of LOTUS, g-factor maps and mean g-factor were calculated using reconstructions that employ Equation 10 for $R_S = 5$ and $R_{IP}=2$. $C_\phi$ from 0 to 1 in 0.1 increments as well as $C_\phi = 0.618$ (inverse of Golden ratio) were used. To quantitatively evaluate quality between CS reconstructions that employed Equation 14, Mean Square Error (MSE), Structural Similarity Index (SSIM) and Entropy vs. $\lambda$ were plotted for each candidate oscillation period $(C_\phi)$. $\lambda$ was varied from 0.01 to 0.063. Based on these results, $C_\phi$ resulting in greatest reduction in error was used for subsequent comparison against other trajectories.

### 4.3.2 Trajectory Comparisons

#### 4.3.2.1 Point Spread Functions

To shed light on the controlled aliasing introduced by each trajectory, point spread functions (PSF) were constructed via PSF = $S^H \mathbf{1}$ where $\mathbf{1}$ represents an input vector of ones and H signifies a conjugate transpose[6] and convolved with numerical phantoms for spiral, CAIPI spiral, and LOTUS acquisitions for $R_S = 5$.

#### 4.3.2.2 G-Factor Maps and Error Metrics

G-factor was calculated for $R_{IP}=1, 2$, and 3 and $R_s = 2$ and 5. To quantitatively evaluate quality between CS reconstructions and the ground truth, MSE, SSIM, and entropy were calculated for $\lambda$ varying from 0.010 to 0.063. Both 2 and 5 simultaneous slices were investigated, with $R_{IP} = 2$. The remaining parameters for all simulations were: FOV=200x200 mm$^2$, in-plane resolution=1.5x1.5 mm$^2$, maximum gradient strength = 30 mT/m, and maximum slew rate = 120 mT/m/msec.

## 4.4 IN VIVO

Two healthy volunteers were scanned on a 3T MRI scanner (Siemens Magnetom Prisma) at Western University's Centre for Functional and Metabolic Mapping (80mT/m maximum gradient strength and 200T/m/s maximum slew rate). This study was approved by the Institutional Review Board at Western University, and informed consent was obtained prior to scanning.

Field monitoring was performed using a $^{19}$F commercial field probe system with 16 probes mounted on a scaffold that optimizes placement (Skope Clip-on Camera) to obtain field dynamics that were fit to second-order spherical harmonics.[11] The field monitoring scans were performed separately from subject scanning using identical sequence parameters. The collected k-space data was coil compressed to 20 virtual coils to improve reconstruction speed,[34–37] and noise correlation between receivers was corrected using pre-whitening before reconstructions.[38]

A Cartesian dual-echo gradient-echo acquisition was used to estimate $B_0$ maps for inclusion in $S$ in a model-based CS reconstruction to correct for static off-resonance effects. The imaging parameters were as follows: FOV = 240 x 240 mm$^2$, spatial resolution = 1.5 mm isotropic, and $TE_1/TE_2$ = 4.08/5.10 ms. Sensitivity coil maps were estimated from the first echo using ESPIRIT.[39]

Diffusion-weighted acquisitions were performed, comparing CAIPI EPI (vendor diffusion MRI sequence), CAIPI Spiral (in-house developed, starting from vendor CAIPI EPI), and LOTUS (in-house developed, starting from vendor CAIPI EPI) trajectories for: (a) $R_S$ = 4, $R_{IP}$ = 2, and TR = 7500ms; and (b) $R_S$=2, $R_{IP}$=4, and TR=15000ms. TR was doubled for $R_S$=2 to keep interslice crosstalk comparable between the two cases. The readout times for CAIPI EPI, CAIPI Spiral, and LOTUS for SMS $R_S$ = 4 and $R_{IP}$ = 2 were 43ms, 47ms, and 46ms respectively. The readout times for CAIPI EPI, CAIPI Spiral, and LOTUS for $R_S$ = 2 and $R_{IP}$ = 4 were 14ms, 19ms, and 19ms respectively. The corresponding gradient trajectories are shown in Supporting Figure S2. One 2D spiral acquisition was conducted as a reference scan with $R_S$ = 1, $R_{IP}$ = 2, TR = 7500ms, and readout time = 47ms. The remaining common parameters were: FOV =218 mm x 218 mm, 1.7mm isotropic resolution, 60 slices, and TE=62ms. The imaging protocol used monopolar pulsed gradient spin-echo encoding using a b-value of b=1000s/mm$^2$ and 6 diffusion directions, plus one b= 0s/mm$^2$.

Separate image reconstruction that employed either Equation 10 (i.e., k-space masking) or Equation 14 (i.e., CS) were performed using the MatMRI toolbox[27,30]. Mean diffusion-weighted images were computed by taking the average over the six diffusion encoding directions. Following the reconstruction, $\hat{g}$ was computed using the pseudo-multiple replica method with reconstructions that employ Equation 10.[28] All image reconstructions were allowed to fully converge (i.e., no early stopping implicit regularization[26] was performed). Mean ($\hat{g}_{mean}$) and peak ($\hat{g}_{peak}$) g-factors were computed in a large region-of-interest manually selected to contain all the brain tissue across all 60 slices. The peak value represents the 95th percentile, indicating that 95% of the g-factor values fall below this threshold. Only the b = 0 scans were used for g-factor calculations. The MRtrix3[40] package was used to estimate the diffusion tensor and obtain FA.

# 5 RESULTS

## 5.1 SIMULATIONS

### 5.1.1 Investigation of $k_z$ Oscillation Period

Figure 2 (a) shows the LOTUS trajectories for $C_\phi$ of 0.1, 0.3, 0.618, and 0.9, (b) shows mean g-factor for all simulated $C_\phi$ values, and (c) shows the g-factor maps for $C_\phi$ of 0.1, 0.3, 0.618, and 0.9. Significant improvements in g-factor are observed for $C_\phi = 0.3$ and 0.618. Figure 2d shows MSE, SSIM, and Entropy plotted against $\lambda$ for each of the oscillation periods depicted in Figure 2a. Overall, maximum reduction in error is observed in all metrics for $C_\phi = 0.618$, which is the inverse of the golden ratio. Thus, this value was used when comparing LOTUS against other trajectories.

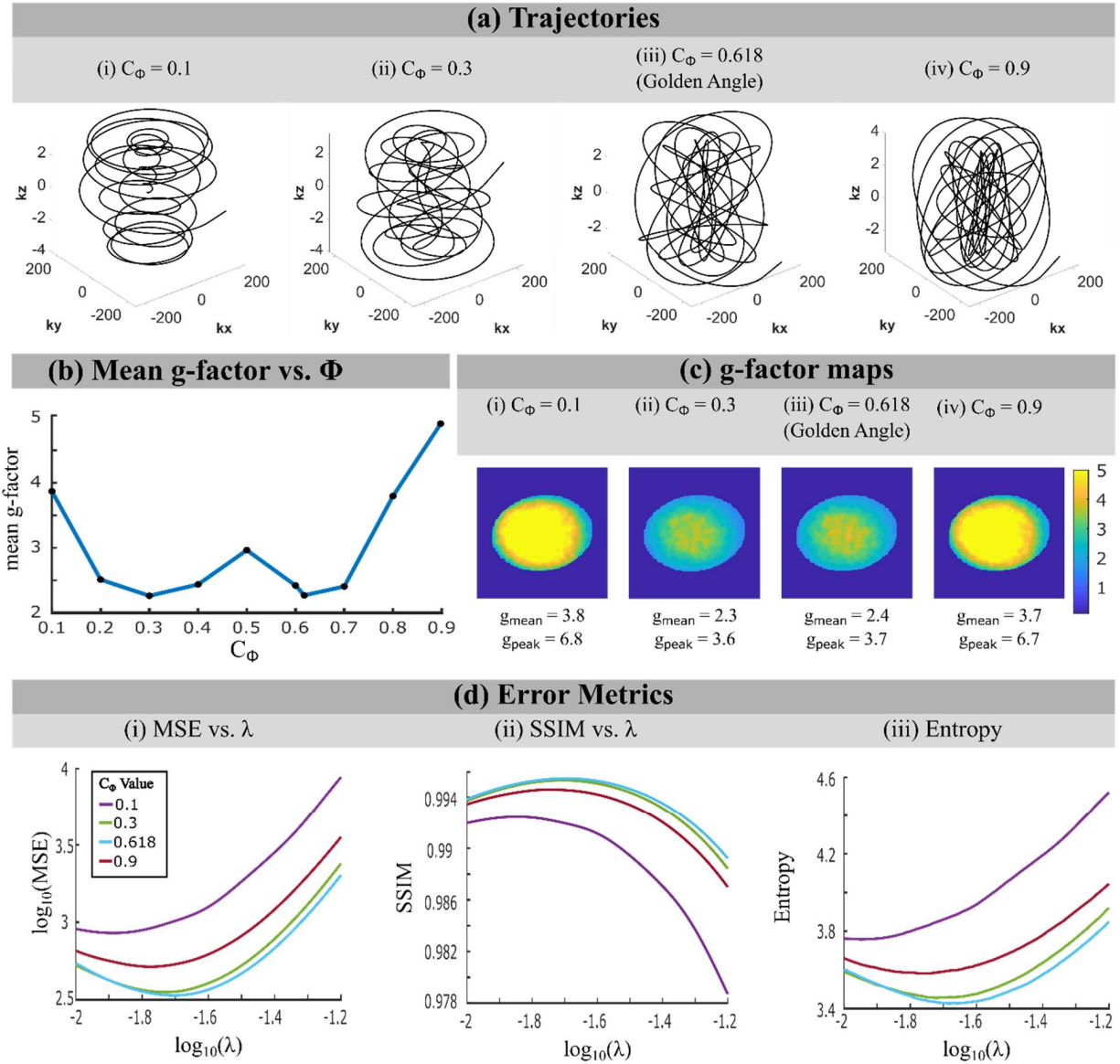

*Figure 2: (a) LOTUS trajectory visualized in 3D for (i) $C_\Phi = 0.1$, (ii) $C_\Phi = 0.3$, (iii) $C_\Phi = 0.618$, (iv) $C_\Phi = 0.9$ (b) Mean g-factor vs. $C_\Phi$ for $C_\Phi$ values ranging from 0.1 to 0.9 in increments of 0.1 (c) g-factor maps of a sample slice for (i) $C_\Phi = 0.1$, (ii) $C_\Phi = 0.3$, (iii) $C_\Phi = 0.618$, (iv) $C_\Phi = 0.9$ (d) Simulations of compressed sensing reconstructions of a numerical phantom, showing (i) Mean Square Error (MSE), (ii) Structural Similarity Index (SSIM), and (iii) Entropy vs. CS tuning parameter $\lambda$ for $C_\Phi = 0.1, 0.3$ $0.618$, and $0.9$. Constant parameters include $R_S = 5$ and $R_{IP} = 2$.*

### 5.1.2 Comparison of Trajectories

*5.1.2.1 Point Spread Functions*

Figure 3a shows the PSF for each simultaneously acquired slice for spiral, CAIPI Spiral, T-Hex, and LOTUS for $R_S$ = 5. PSFs show more incoherent aliasing for LOTUS compared to the other trajectories. Increased incoherent aliasing is observed in outer slices for all trajectories but spiral due to the larger field offsets from the z-gradients applied for controlled aliasing. Figure 3b shows the PSF from each slice in Figure 3a convolved with a numerical phantom. Note that the image acquired from simple gridding of the data with no parallel imaging reconstruction is the sum of the images in Figure 3a across all slices, and controlled aliasing causes overlapping signal to be distributed through the FOV.

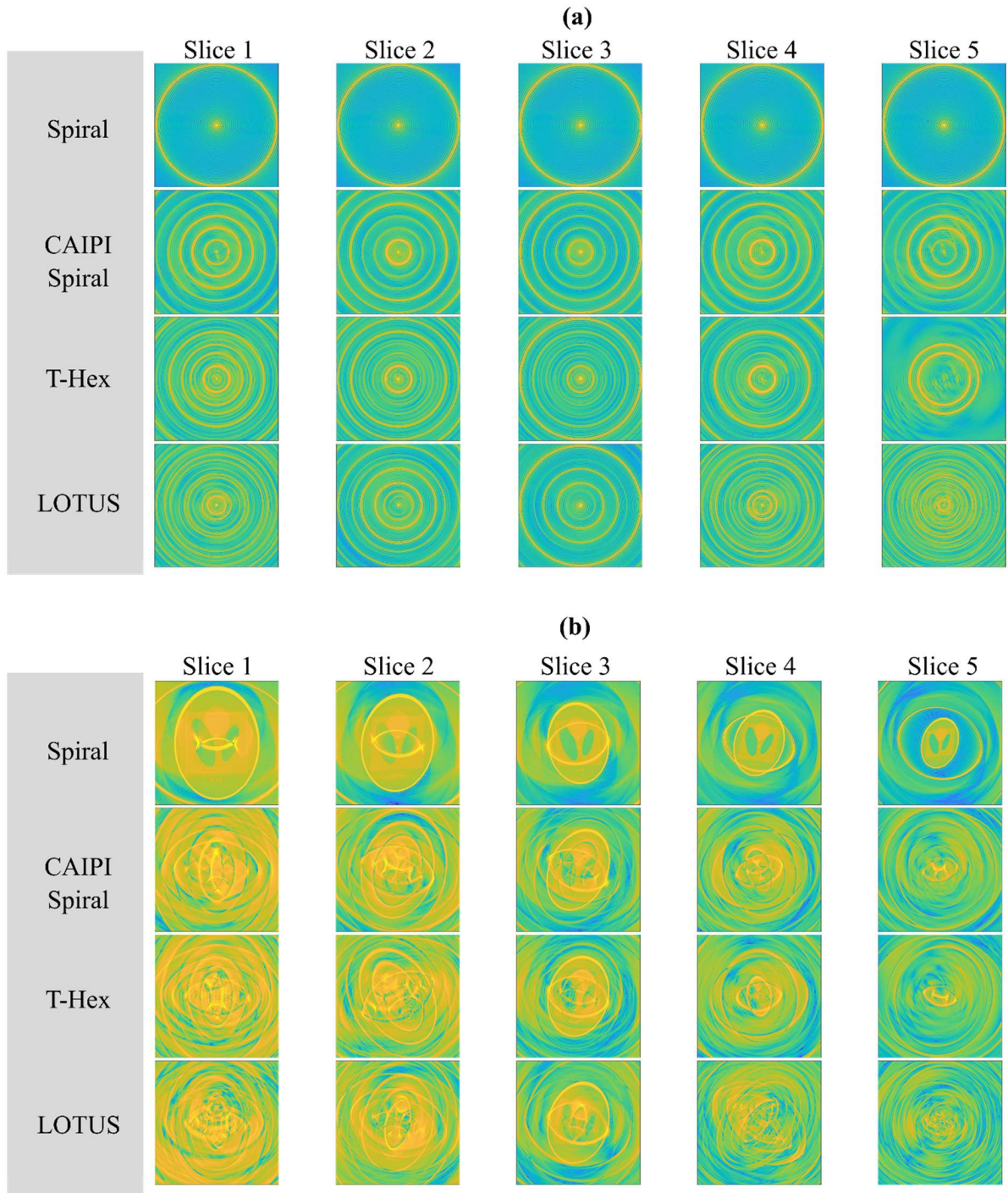

*Figure 3: (a) Point spread functions (PSF) comparing spiral, CAIPI spiral, and LOTUS acquisition for $R_S=5$; (b) Corresponding PSFs convolved with numerical phantom. Note that different slices were created by applying affine transformations to the standard Shepp-Logan phantom.*

*5.1.2.2 G-Factor Maps*

Figure 4 shows g-Factor maps for $R_S$ = 2 and $R_S$ = 5. The center slice is shown as a sample for $R_{IP}$ = 1, 2, and 3. Mean g-factor (over all slices) is shown below each computed image. G-factors were comparable for $R_S$ =2, but significant improvement is seen using 3D non-Cartesian trajectories for $R_S$ = 5. LOTUS exhibits dramatically reduced g-factors for $R_S$ =5.

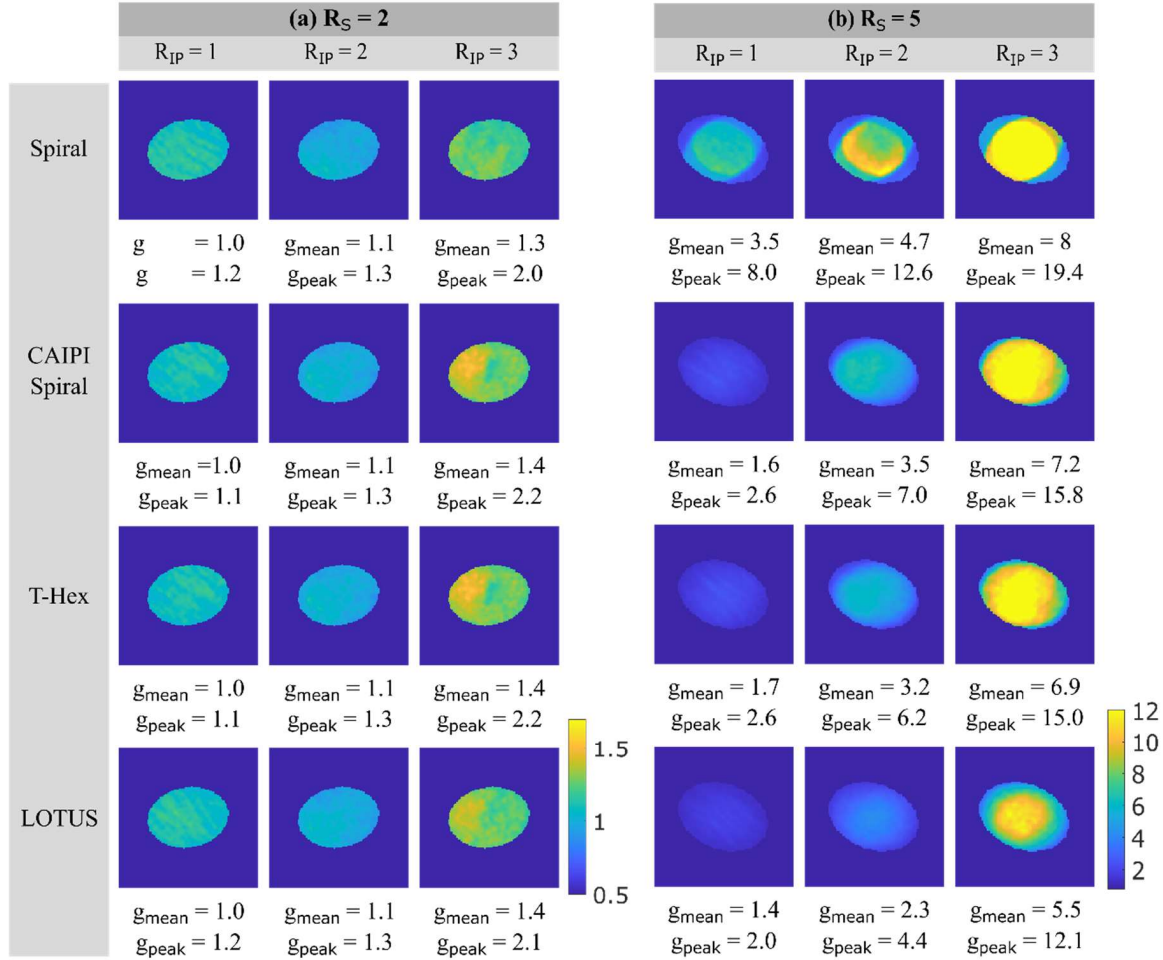

*Figure 4: g-factor maps for a sample slice of a numerical phantom for Spiral, CAIPI Spiral, and LOTUS trajectories. In-plane undersampling rates ($R_{IP}$) of 1, 2, and 3 for each of (a) $R_S$ =2 and (b) $R_S$ =5 were investigated. Large improvements for LOTUS over the other trajectories are observed with many simultaneous slices. Mean g-factor ($g_{mean}$) and peak g-factor ($g_{peak}$) across all simultaneous slices are reported for each case.*

*5.1.2.3   Compressed Sensing Image Reconstructions*

Figure 5 compares spiral, CAIPI Spiral, T-Hex, and LOTUS trajectory performance using MSE, SSIM, and entropy for both $R_S = 2$ and $R_S = 5$, both with $R_{IP} = 2$. All error metrics showed comparable results for $R_S = 2$ between trajectories. However, LOTUS performed best in all metrics with increased slice acceleration. For $R_S = 5$, coherent artefacts and blurring near edges in the phantom can be observed, which are mostly benign with LOTUS (Figure 5e).

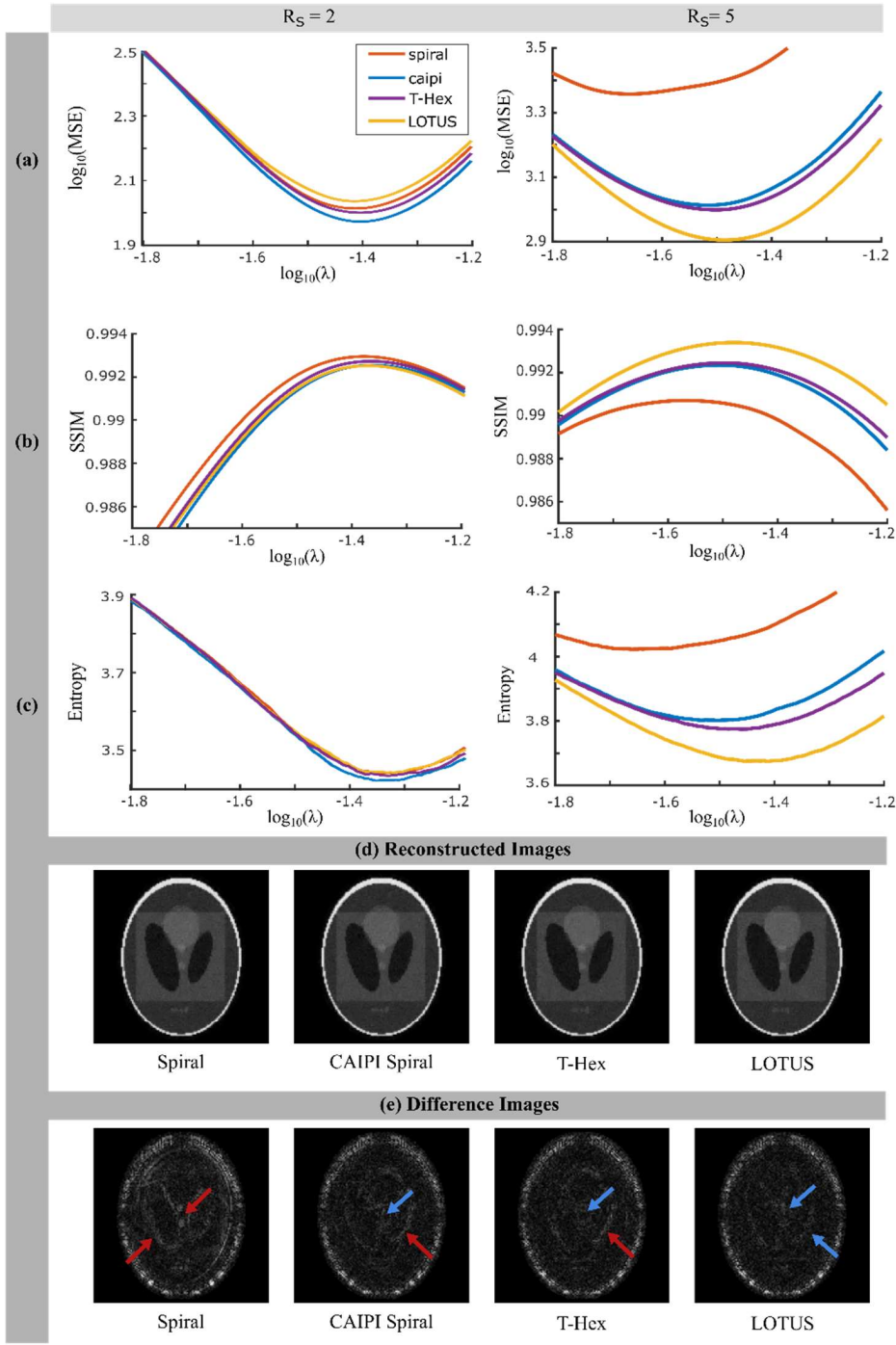

*Figure 5: Simulations of compressed sensing reconstructions of a numerical phantom, showing (a) Mean Square Error (MSE), (b) Structural Similarity Index (SSIM), and (c) Entropy vs. CS tuning parameter λ for $R_S = 2$ and $R_S = 5$ and an in-plane undersampling rate of 2 (d) Images reconstructed using λ value that minimizes MSE for $R_S = 5$ (e) Corresponding difference images showing areas with more artefact (red arrows) and areas where image features are better preserved (blue arrows)*

## 5.2 IN VIVO

Figure 6 shows mean diffusion-weighted images (reconstructed with k-space masking, Equation 10), averaged over the six diffusion encoding directions, and corresponding approximated g-factor ($\hat{g}$) maps (using Equation 10 for reconstructions used in pseudo-multiple replica method) for each candidate trajectory at different rates of SMS and in-plane undersampling. The qualitative differences between trajectories were subtle in the diffusion-weighted images. The non-Cartesian trajectories look slightly blurred compared to EPI due to the broader T2* point spread function associated with spiral.[6] Meanwhile, both 3D non-Cartesian trajectories, CAIPI Spiral and LOTUS, exhibit large g-factor advantages compared to CAIPI EPI. It is difficult to discern differences in performance between CAIPI spiral and LOTUS for $R_S$ = 2 and $R_{IP}$ = 4; however, g-factor advantages were evident for LOTUS or $R_S$ = 4, with a mean value 11% lower than CAIPI spiral. These observations suggesting that that the benefits of LOTUS are less evident with increased in-plane acceleration, but more evident with increased slice acceleration, were similar to the simulations. Similar results were observed in both subjects.

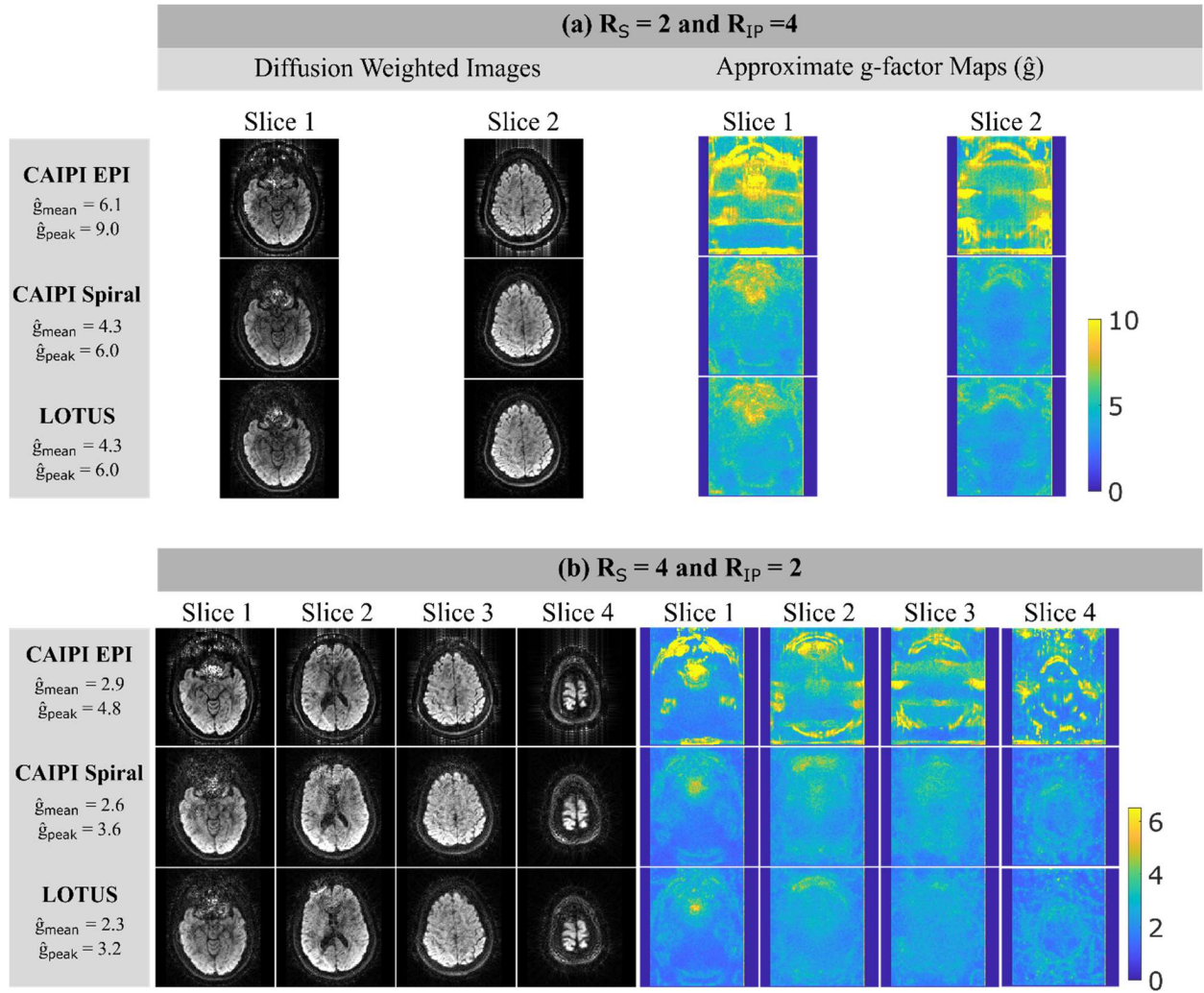

*Figure 6: Diffusion-weighted images (Equation 10) of simultaneously acquired slices and corresponding approximated g-factor maps (Equation 13) for prospectively acquired in-vivo scans for (a) $R_S=2$ and $R_{IP}=4$ and (b) $R_S=4$ and $R_{IP}=2$. For each acquisition, $\hat{g}_{mean}$ and $\hat{g}_{peak}$ were calculated across all 60 slices.*

Figure 7 shows the FA maps for the mean diffusion-weighted images averaged over the six diffusion encoding directions that were computed using Equation 10 for EPI CAIPI, CAIPI Spiral, and LOTUS acquisitions. Comparison of the FA maps revealed that LOTUS acquisitions demonstrate reduced noise in comparison to the EPI CAIPI and CAIPI spiral acquisitions. These trends were especially visible for Subject 1 whose brain filled more of the FOV, which is expected because smaller objects will have smaller g-factor noise amplification due to more effective controlled aliasing. Benefits were more evident with increased slice acceleration.

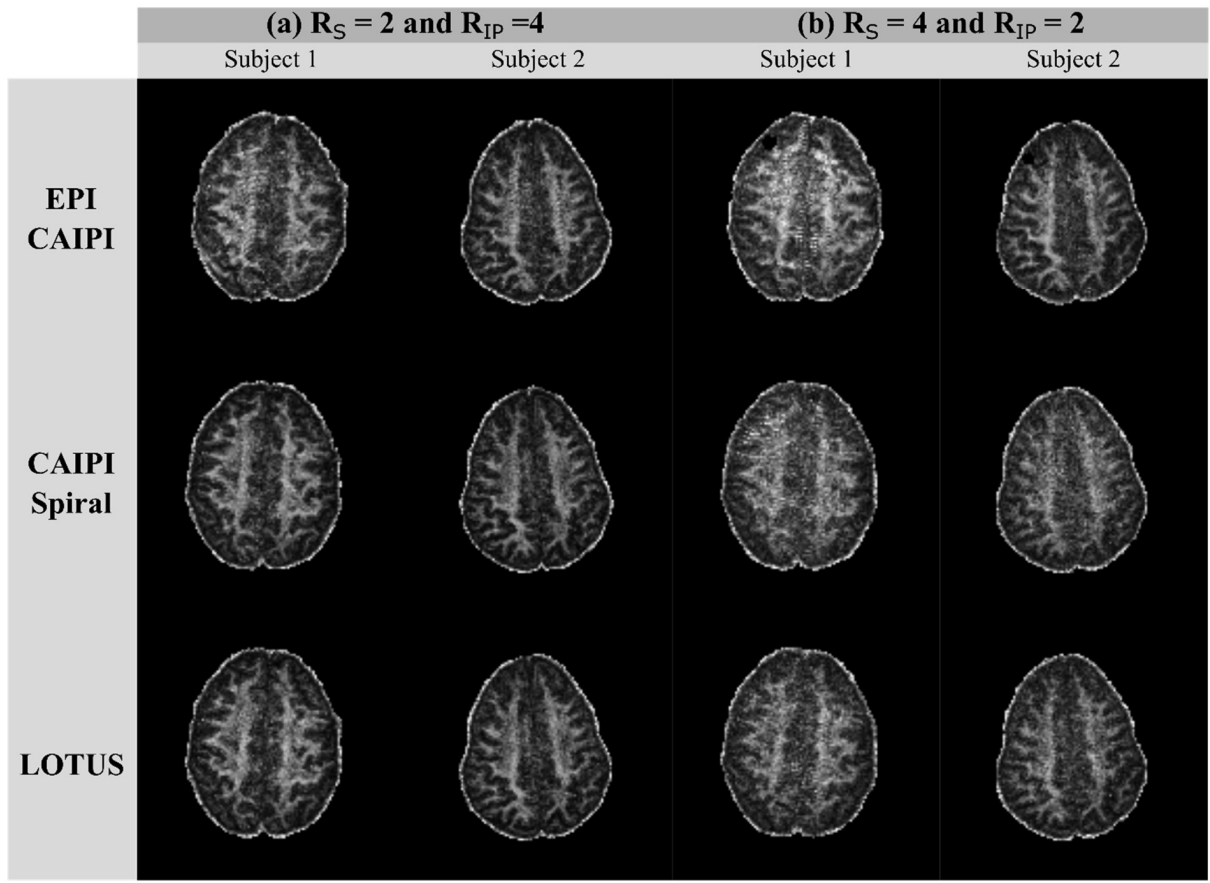

*Figure 7: Sample slice of FA Maps determined using Equation 10 for two subjects. Images were acquired using EPI CAIPI, CAIPI Spiral, and LOTUS for (a) $R_S$=2 and $R_{IP}$=4 and (b) $R_S$=4 and $R_{IP}$=2.*

Figure 8 shows the FA maps for the mean diffusion-weighted images averaged over the six diffusion encoding directions that were computed using CS reconstructions for EPI CAIPI, CAIPI Spiral, and LOTUS acquisitions. Comparison of the FA maps revealed that LOTUS acquisitions demonstrate reduced artefact in comparison to the EPI CAIPI and CAIPI spiral acquisitions. Again, benefits were more evident with increased slice acceleration and larger brain size.

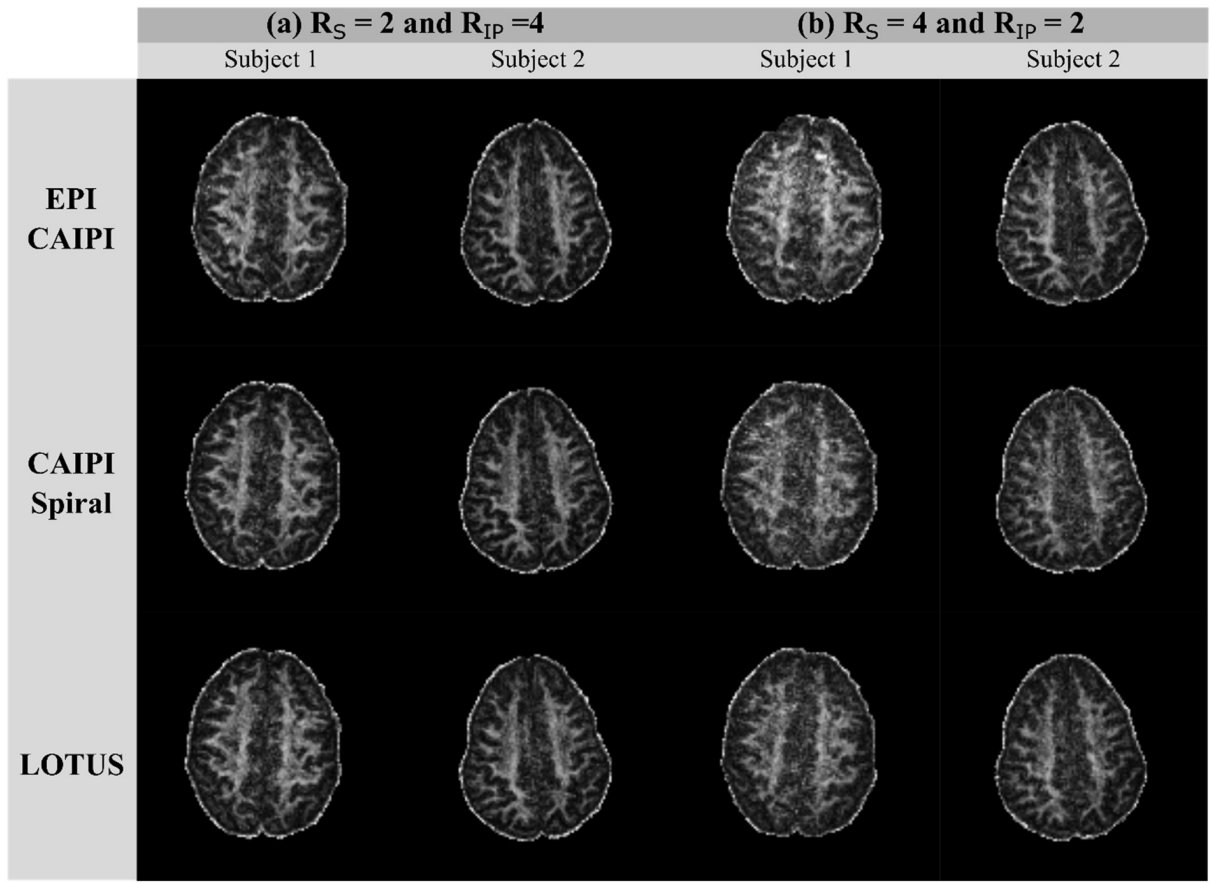

*Figure 8: Sample slice of FA Maps (determined using a compressed sensing reconstruction employing equation 14) for two subjects. Images were acquired using EPI CAIPI, CAIPI Spiral, and LOTUS for (a) $R_S$=2 and $R_{IP}$=4 and (b) $R_S$=4 and $R_{IP}$=2.*

# 6 DISCUSSION

Simulated and *in-vivo* results both indicate benefits of LOTUS for high rates of slice acceleration, which can likely be attributed to the controlled aliasing varying more between slices. However, for low rates of slice acceleration all trajectories had comparable performance. MSE was marginally larger for LOTUS in simulated results for $R_S = 2$, but this did not translate to a lower SSIM, making no trajectory a clear optimum choice for low SMS rates.

Incoherence in the sampling pattern, or "noise-like" artefact in the sparsifying transform domain, is also essential for CS efficacy. Theoretically, CS would be most effective with a completely random subset of k-space, providing low coherence. However, since this is not practical due to hardware and physiological constraints, sampling trajectories are designed to follow relatively smooth lines and curves. 2D EPI acquisitions are limited because they require full sampling in the readout direction. Thus, the design of incoherent sampling patterns is limited to only one degree of freedom, along the phase encoding direction. In 2D non-Cartesian acquisition, undersampling is distributed across x and y which increases incoherence, and in 3D non-Cartesian acquisitions, there are 3 degrees of freedom to design incoherent sampling patterns. LOTUS performed better than CAIPI spiral in CS recons, which could be due to greater incoherence enforced in 3D. However, it is generally difficult to discern if improved CS performance is due to this incoherence or the reduction in g-factor, since these two sampling features are strongly linked. That said, the changes with $C_\phi$ provide evidence that the pseudo-random sampling provides a benefit for CS beyond g-factor, given that even though g-factor is comparable for $C_\phi$ of 0.3 and 0.618, the value of 0.618 provides slightly lower MSE in CS reconstructions (Figure 2). This can likely be attributed to the increased incoherence in the z-direction. To investigate this hypothesis further, PSFs were computed for LOTUS for $C_\phi$ $of$ 0.3 $and$ 0.618. These were compared to other candidate trajectories by computing their entropy (Supporting Figure S3). Pure random sampling would result in a PSF with maximal entropy, and increasing coherence (e.g., via high amplitude rings observable in many of the PSFs in Figure 3) will decrease the entropy. Here, highest entropy is observed for LOTUS with $C_\phi = 0.618$, thus supporting the supposition that a value of $C_\phi$ based on the golden ratio maximizes incoherence, which improves g-factor and may modestly improve CS performance.

Much larger g-factor noise amplification is observed in the in-vivo EPI acquisitions, which is expected because in-plane undersampling only occurs in one direction compared to the undersampling being distributed across x and y for spiral.[5] The distribution of undersampling in 3D further improves noise amplification for CAIPI Spiral and LOTUS. It should be noted that the large noise amplification was not obviously evident in the raw in vivo images because a high SNR acquisition with large slice thickness was used to maximize the quality of the g-factor computations, which was the primary concern for the experiments. Nevertheless, qualitative differences were still evident in FA even at this high SNR.

LOTUS has been formulated synergistically with the spiral gradient equations and requires no other parameters to be set other than $C_\phi$, which our results suggest can be fixed to a value of 0.618. Blipped approaches, on the other hand, require setting blip durations and slew rates. We have found practical challenges with blipped approaches, because blips overlap near the center of the spiral if blips are too long in duration (resulting in trajectory generation failure), but readout duration is lengthened if too many gradient resources are devoted to them to make them shorter (i.e., devoting net slew to blips reduces the slew available to turn the spiral in-plane). As a result, we needed to tune blip parameters for each trajectory, which was not required for LOTUS. It is also worth noting that LOTUS is also compatible with alternative spiral-like approaches, such as variable density spiral, requiring no changes from the formulation introduced here. Finally, we have also found that LOTUS generally has very little impact on readout duration. For example, at a slice acceleration factor of $R_S$=5, the readout duration was 8.57 ms for LOTUS and 8.54 ms for the conventional spiral, a negligible difference of just 0.03 ms.

# 7  CONCLUSION

In this work, we introduced a 3D non-Cartesian sampling trajectory, referred to as "Laterally Oscillating Trajectory for Undersampling of Slices" (LOTUS). Trajectory performance was compared to other trajectories in diffusion MRI at 3T, and g-factors were assessed. Simulations and *in vivo* acquisitions indicated benefits of LOTUS compared to blipped approaches, particularly for high rates of slice acceleration. By enabling higher rates of slice acceleration, LOTUS shows promise for decreasing scan time, which is especially beneficial for diffusion MRI.

# 8  ACKNOWLEDGEMENTS

This work was supported by NSERC Discovery Grants (RGPIN-2025-05597) and the Ontario Graduate Scholarship program. The authors thank Mr. Trevor Szekeres and David Reese for aiding with data acquisition.

# 9  DATA AVAILABILITY STATEMENT

Code for LOTUS trajectory generation and expanded encoding CS image reconstructions is available at https://gitlab.com/cfmm/matlab/matmri.

# 11 Supporting Figures

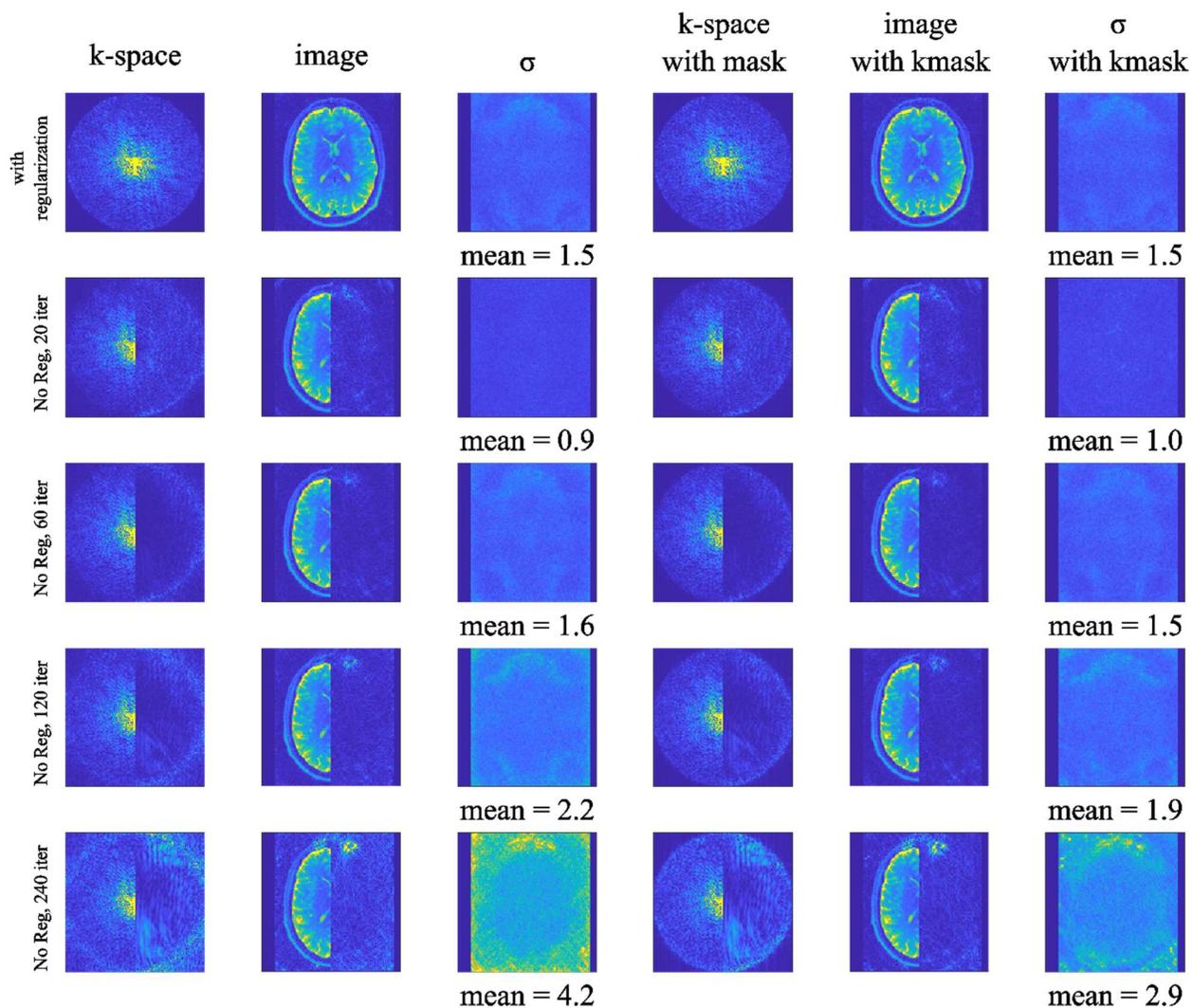

Supporting Figure S1: From the top row, the proposed method to explicitly enforce no k-space signal in unacquired corner regions (Equation 10) enables full convergence of conjugate gradient iterations without excessive noise amplification. Rows 2 to 5 show that with the standard approach (Equation 9), noise amplification increases with increasing numbers of conjugate gradient iterations. The right half of k-space and object-domain images show the difference from the top row (scaled by 2x), and σ is the noise standard deviation determined using the pseudo-multiple replica method. Notably, masking the corners of k-space after finishing iterations (columns 4 to 6) does not fully account for this noise amplification, which also occurs within the circular region of acquired k-space. Even though the noise amplification originates from the corners of k-space, the noise amplification that also occurs within the acquired region of k-space may occur because the algorithm will inherently seek a solution without sharp edges in k-space (since this would cause rippling in the object domain that does not agree with the smooth receiver profiles and true underlying image).

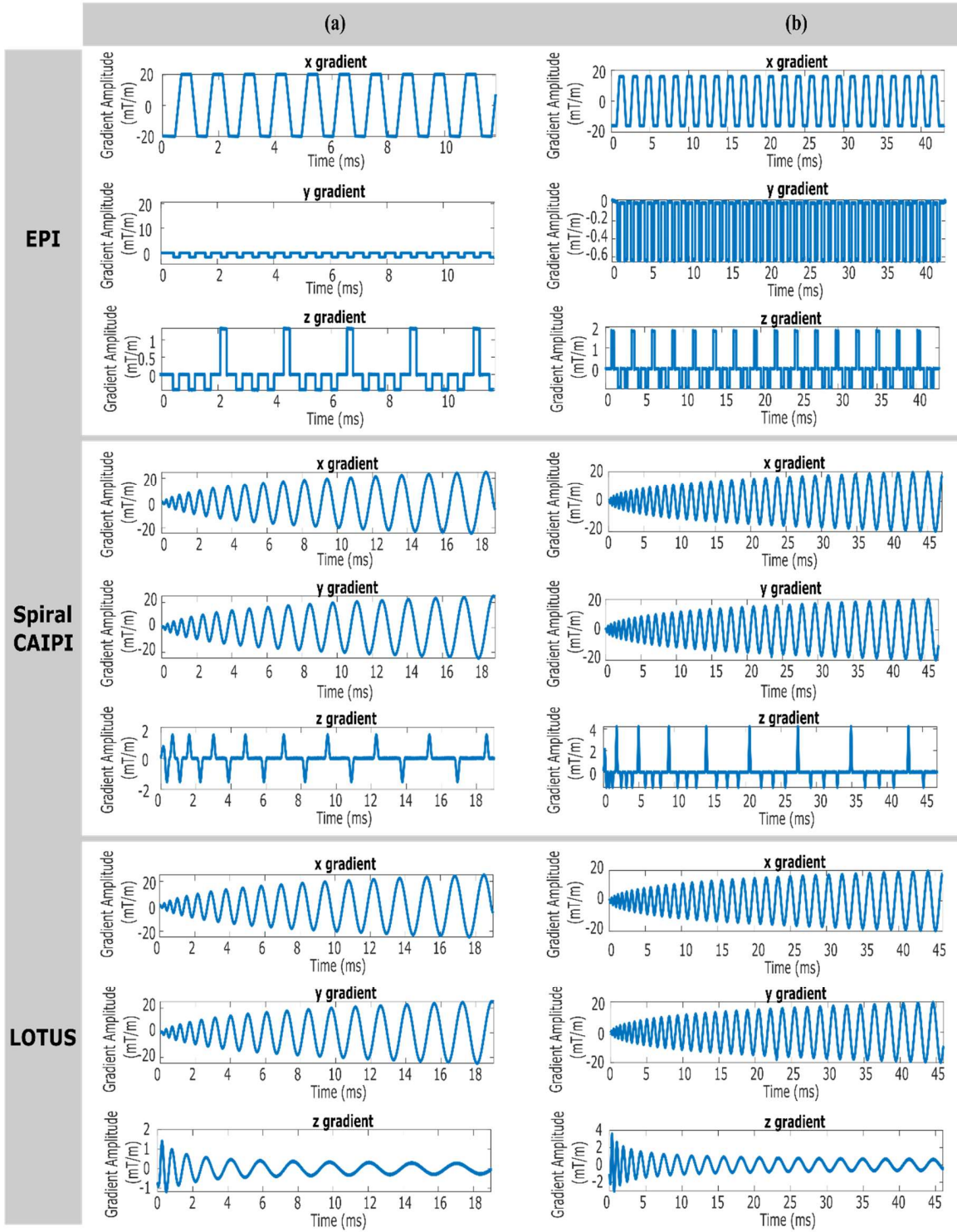

Supporting Figure S2: Gradient trajectories for EPI CAIPI, Spiral CAIPI, and LOTUS scans conducted for (a) $R_S$=2 and $R_{IP}$=4 and (b) $R_S$=4 and $R_{IP}$=2, measured using field monitoring.

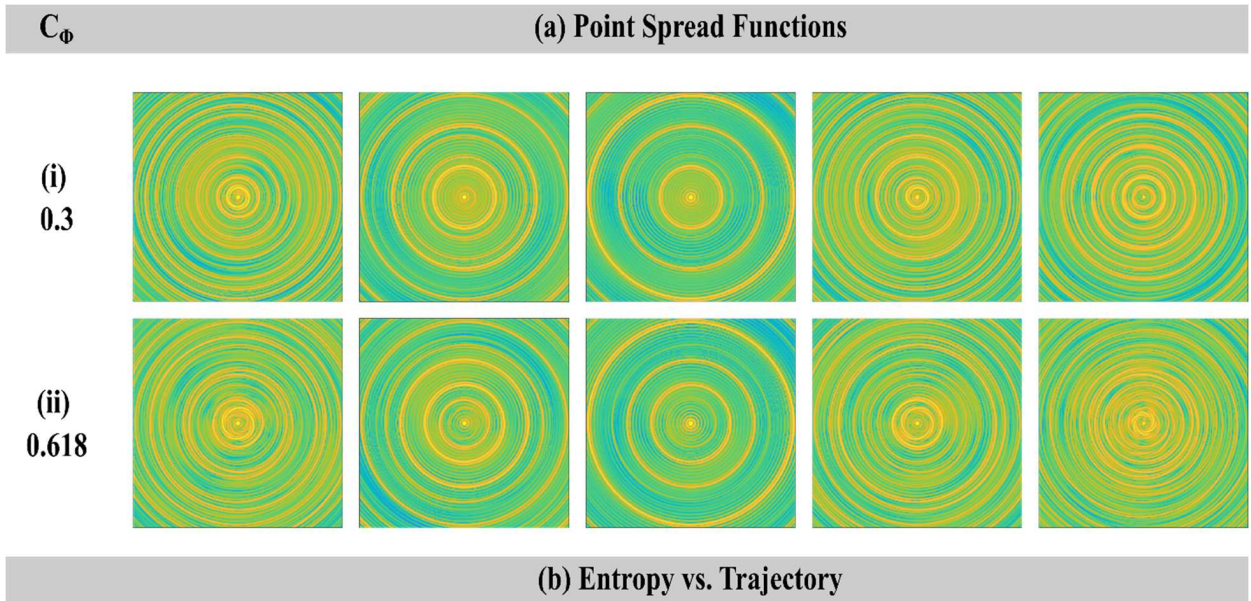

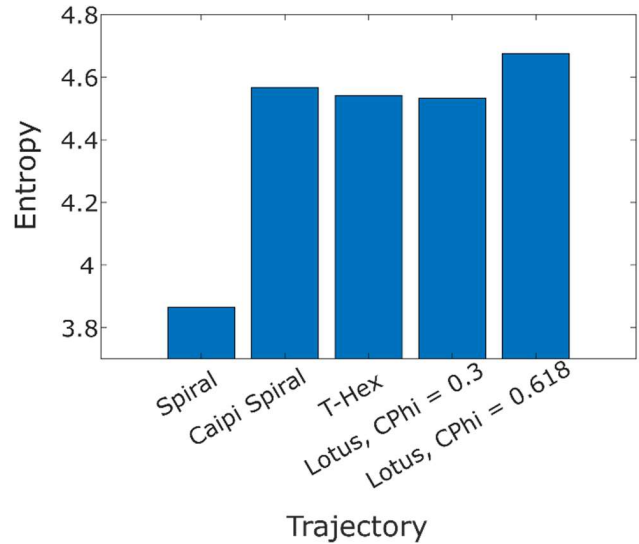

Supporting Figure S3: (a) Point spread functions (PSF) comparing LOTUS acquisition where $R_S=5$ for (i) $C_\Phi = 0.3$ and (ii) $C_\Phi = 0.618$, and (b) histogram comparing entropy for PSFs comparing spiral, CAIPI spiral, T-Hex and two LOTUS acquisitions ($C_\Phi = 0.3\ and\ 0.618$) for $R_S=5$.